%% file: paper.tex
\documentclass{juliacon}
\usepackage{balance}

\usepackage{hyperref}
\hypersetup{
    colorlinks=true,
    linkcolor=blue,
    citecolor=blue,
    urlcolor=blue,
    filecolor=blue,
    pdfborder={0 0 0},
    breaklinks=true
}

\begin{document}

\input{header}

\maketitle

{\let \thefootnote \relax \footnotetext{*Email: \texttt{adriano.meligrana@diag.uniroma1.it}}}

\begin{abstract}
StreamSampling.jl is a Julia library designed to provide general and efficient methods for sampling from data streams in a single pass, even when the total number of items is unknown. In this paper, we describe the capabilities of the library and its advantages over traditional sampling procedures, such as maintaining a small, constant memory footprint and avoiding the need to fully materialize the stream in memory. Furthermore, we provide empirical benchmarks comparing online sampling methods against standard approaches, demonstrating performance and memory improvements.
\end{abstract}

\section{Introduction}

Random sampling from data streams is a fundamental operation in data analysis, particularly when dealing with massive datasets that exceed available memory or continuous streams of indeterminate length. The methods covered here fall under the umbrella of \textit{online sampling}, i.e. algorithms that process stream elements sequentially in a single pass, which can be broadly divided into two subcategories: reservoir sampling and sequential sampling \cite{vitter1985random, vitter1987efficient}. 

Reservoir sampling algorithms maintain a sample of a fixed size $K$, which is dynamically updated as stream elements are processed. They guarantee that, at any point during the process, the collected sample is representative of the portion of the stream seen so far. This makes them ideal for unbounded, continuous streams where the total population size $N$ is unknown \cite{vitter1985random}. 

Sequential sampling algorithms, by contrast, are typically used when either the total number of elements in the stream ($N$) or the exact total weight ($W_N$) is known in advance \cite{vitter1987efficient, shekelyan2021sequential}. These methods return an ordered sample of the stream, without keeping track of previous sampled elements. A sequential algorithm can compute, from a single random variate, how many elements to skip over before the next selection, without the need to keep a reservoir. This can make sequential methods more efficient than reservoir methods when prior knowledge of $N$ or $W_N$ is available. \texttt{StreamSampling.jl} provides a comprehensive native Julia suite of algorithms covering both categories. However, a fundamental advantage of reservoir methods over sequential methods is that they maintain a representative sample in memory at all times, even when only a portion of the $N$ elements have been processed, a property that sequential methods lack by design.

\section{Statement of Need}

Online sampling techniques have been implemented across a range of programming languages and frameworks, typically addressing specific use cases rather than providing a unified suite of algorithms. For instance, eBay's \texttt{tsv-utils} provide the \texttt{tsv-sample} command-line tool, which supports simple random sampling and weighted reservoir sampling over tabular streams, as well as Bernoulli sampling and distinct sampling for streaming scenarios. In the POSIX environment, the GNU Coreutils \texttt{shuf} binary \cite{gnucoreutilsshuf} implements unweighted reservoir sampling when the \texttt{-n} option is used to extract a fixed number of lines from potentially unbounded inputs, without needing to know the total line count in advance.

In the Big Data domain, Apache DataFu \cite{sumbaly2013datafu} provides User-Defined Functions (UDFs) for Apache Pig, including both unweighted (\texttt{ReservoirSample}) and weighted (\texttt{WeightedReservoirSample}) reservoir sampling routines. These implementations are designed for the Hadoop/Pig ecosystem and are not directly usable outside of it. Similarly, in Python, the River library \cite{montiel2021river} focuses on online machine learning and includes sampling utilities primarily targeted at handling class imbalance in streaming classification and regression tasks, rather than general-purpose random sampling. Finally, the Clojure library \texttt{bigmlcom/sampling} most closely resembles \texttt{StreamSampling.jl} in terms of scope. However, it does not support the full range of reservoir and sequential sampling variations offered by \texttt{StreamSampling.jl}. Furthermore, the project appears to be inactive, with its last commit dating back seven years.

Compared to these alternatives, \texttt{StreamSampling.jl} provides a comprehensive suite of sampling methodologies directly within the Julia programming language \cite{bezanson2017julia}, covering both reservoir and sequential paradigms, with and without replacement, and with or without weights. This makes it the only package, to the author's knowledge, that offers such breadth in a single, self-contained library. Furthermore, its design follows Julia's standard iterable protocol and is compatible with the \texttt{OnlineStats.jl} \cite{day2020onlinestats} API, enabling excellent integration with existing Julia data pipelines.

\section{Research Impact Statement}

Although \texttt{StreamSampling.jl} is a recent package, we believe the work already shows credible near-term scholarly significance. First, the benchmark results reported in Section~\ref{sec:benchmarks} show that stream-native sampling can reduce both runtime and memory consumption relative to \texttt{StatsBase.sample} by avoiding full materialization of the population. These gains make iterator-based sampling practical in settings where population-based methods incur substantial allocation overhead.

Second, the out-of-core experiment in Section~\ref{sec:applications} demonstrates usefulness beyond synthetic iterator benchmarks. On a 100 GB Arrow dataset, the stream-based methods remain feasible at scales where the chunked \texttt{StatsBase.sample} baseline is slower and eventually fails due to out-of-memory errors.

Third, the package serves as a software vehicle for recent research results, since its implementation of \texttt{AlgWRSWRSKIP} is tied to a recent publication in the field \cite{meligrana2026weighted}.

Taken together, these results provide evidence that the library already transfers current data-stream sampling research into practical Julia workflows and is positioned for near-term use in statistical computing and large-scale data analysis.

\section{Implemented Methods}

\texttt{StreamSampling.jl} categorizes its implemented methods along three main axes: the underlying logic (reservoir vs. sequential), the sampling scheme (with vs. without replacement), and the inclusion probabilities (weighted vs. unweighted).

For reservoir sampling without replacement, \texttt{AlgR} and \texttt{AlgL} \cite{vitter1985random} are provided, whereas \texttt{AlgRSWRSKIP} \cite{park2004reservoir} covers the with-replacement counterpart. For weighted reservoir sampling, \texttt{AlgARes} and \texttt{AlgAExpJ} \cite{efraimidis2006weighted} handle the without-replacement case, and \texttt{AlgWRSWRSKIP} \cite{meligrana2026weighted} is used for the with-replacement case.
In the sequential sampling domain, methods like \texttt{AlgD} \cite{vitter1987efficient} and \texttt{AlgHiddenShuffle} \cite{shekelyan2021sequential} are employed for unweighted without-replacement sampling. \texttt{AlgORDSWR} \cite{bentley1980generating} is used for sequential sampling with replacement, and \texttt{AlgORDWSWR} \cite{startek2016asymptotically} for weighted sequential sampling with replacement.

Notably absent from this set is a weighted sequential method without replacement. This is a fundamental mathematical impossibility when the only prior information over the unobserved elements is the total remaining weight of the stream $W_N$, which is what it is usually assumed to be known for weighted sequential sampling algorithms. In sampling with replacement, every draw is independent and the inclusion probability of any item, that is, the probability that it will appear in the final sample is proportional to its weight over $W_N$, so $W_N$ alone is sufficient. In sampling without replacement, however, selecting an item alters the inclusion probabilities of all subsequent items. To correctly compute the inclusion probability of the next item, one needs to know the individual weights of all remaining elements in the stream, not merely their sum: different configurations of individual weights can yield the same total $W_N$ but require different inclusion probabilities. Knowing only $W_N$ therefore leaves the per-element inclusion probabilities undetermined, making it impossible to draw a correctly weighted sample without replacement in a sequential single pass.

\vspace{-10pt}
\begin{table}[h]
\centering
\resizebox{\columnwidth}{!}{%
\begin{tabular}{|l|c|c|c|c|c|c|}
\hline
\textbf{Method} & \textbf{Ref.} & \textbf{Type} & \textbf{Replacement} & \textbf{Weighted} & \textbf{Time} & \textbf{Space} \\
\hline
AlgR & \cite{vitter1985random} & Reservoir & Without & False & $\mathcal{O}(N)$ & $\mathcal{O}(K)$ \\
AlgL & \cite{vitter1985random} & Reservoir & Without & False & $\mathcal{O}(K\log(N/K))$ & $\mathcal{O}(K)$ \\
AlgRSWRSKIP & \cite{park2004reservoir} & Reservoir & With & False & $\mathcal{O}(K \log{N})$ & $\mathcal{O}(K)$ \\
AlgARes & \cite{efraimidis2006weighted} & Reservoir & Without & True & $\mathcal{O}(N)$ & $\mathcal{O}(K)$ \\
AlgAExpJ & \cite{efraimidis2006weighted} & Reservoir & Without & True & $\mathcal{O}(K \log(N/K))$* & $\mathcal{O}(K)$ \\
AlgWRSWRSKIP& \cite{meligrana2026weighted} & Reservoir & With & True & $\mathcal{O}(K \log W_N)$ & $\mathcal{O}(K)$ \\
AlgD & \cite{vitter1987efficient} & Sequential & Without & False & $\mathcal{O}(K)$ & $\mathcal{O}(1)$ \\
AlgHiddenShuffle& \cite{shekelyan2021sequential} & Sequential & Without & False & $\mathcal{O}(K)$ & $\mathcal{O}(1)$ \\
AlgORDSWR & \cite{bentley1980generating} & Sequential & With & False & $\mathcal{O}(K)$ & $\mathcal{O}(1)$ \\
AlgORDWSWR & \cite{startek2016asymptotically} & Sequential & With & True & $\mathcal{O}(K)$ & $\mathcal{O}(1)$ \\
\hline
\end{tabular}%
}
\caption{Sampling algorithms implemented in \texttt{StreamSampling.jl} ($N$=population size, $K$=sample size, $W_N$=total weight). $^*$The expected $\mathcal{O}(K \log(N/K))$ complexity for \texttt{AlgAExpJ} holds when weights are independent and identically distributed (i.i.d.) draws \cite{efraimidis2006weighted}.}
\label{tab:methods}
\end{table}
\vspace{-5pt}

Table \ref{tab:methods} summarizes the algorithms implemented in the package along with their time and space complexities.

It is important to clarify that the \textit{Time} complexity metric strictly evaluates the running time in terms of the number of random variates that must be generated to obtain a sample of size $K$ from a (possibly weighted) population of size $N$.

\section{Software Design}

The software design has two goals: integration with existing Julia streaming workflows and efficient support for the different sampling algorithms implemented by the package. For reservoir samplers, this led to an interface aligned with \texttt{OnlineStats.jl} \cite{day2020onlinestats}. Rather than introducing a package-specific stateful API, reservoir samplers support \texttt{fit!}, \texttt{value}, and \texttt{merge!}. This makes them composable with existing online and streaming workflows. The use of \texttt{fit!} and \texttt{value} for reservoir sampling has previously been proposed in \cite{efraimidis2015weighted}, while mergeable reservoirs have been discussed in \cite{cormode2011}. The latter operation is particularly important for parallel and distributed sampling workflows.

Reservoir samplers naturally fit the \texttt{OnlineStats.jl} model because they maintain a state that is updated as observations arrive and can later be queried or merged. Sequential samplers, however, have a different computational structure and therefore follow Julia's iterator protocol instead.

The implementation also balances performance with flexibility by providing most reservoir samplers in both mutable and immutable forms. These variants are generated from a shared definition using \texttt{HybridStructs.jl}. The mutable variants are easier to work with for workflows in which surrounding code requires storing references of the sampler object. Empirical testing showed that the immutable variants are generally more performant, and they are therefore retained for cases where more speed is required.

The internal data structures to store the sample are chosen to match the operations required by each algorithm. Weighted sampling without replacement, as implemented by \texttt{AlgARes} and \texttt{AlgAExpJ}, repeatedly compares candidate priorities with the current minimum retained priority to choose if an element needs to be sampled. For this reason, the sample is stored in a binary heap, as is usually recommended \cite{efraimidis2006weighted, efraimidis2015weighted}. This allows the current top-$K$ priorities to be maintained efficiently and also simplifies merging, at the cost of the ordering overhead introduced by the heap. In contrast, the unweighted algorithms and the weighted algorithms with replacement maintain fixed sample slots in flat arrays. This simpler representation allows these algorithms to outperform the heap-based implementations, as shown in Figure~\ref{fig:bench_iter}.

\subsection{Package Interface}

The interface of \texttt{StreamSampling.jl} relies on two main samplers, each made for its sampling domain: \texttt{ReservoirSampler} and \texttt{SequentialSampler}. The package also provides \texttt{itsample}, which, similarly to \texttt{StatsBase.sample}, returns an \texttt{Array}, but, by using stream methods, it can be applied to any iterator.

\subsubsection{Reservoir Samplers}

As previously described, the \texttt{ReservoirSampler} API exposes three main core operations:

\begin{itemize}
    \item \texttt{fit!(sampler, item, [weight])}: Processes a new element from the stream and updates the sample accordingly.
    \item \texttt{value(sampler)}: Returns the current in-memory sample.
    \item \texttt{merge!(sampler1, sampler2, ...)}: Combines multiple reservoir samplers, each typically operating on a different partition of the stream, into a single statistically consistent reservoir.
\end{itemize}

\subsubsection{Sequential Samplers}

Instead of maintaining an in-memory reservoir, \texttt{SequentialSampler} wraps an input iterable together with the required population size (or total weight) and conforms to Julia's standard \texttt{iterate} protocol. Rather than processing each element via \texttt{fit!}, the sampler computes skip lengths on the fly and emits selected elements directly as the iterator is consumed, without any intermediate collection.

For parallel workflows, \texttt{SequentialSampler} provides a \texttt{combine(samples, weights)} function. Similarly to \texttt{merge!}, \texttt{combine} merges samples taken from different partitions into a single globally correct sample by re-weighting each local sample proportionally to its partition's share of the total weight, ensuring that the final output has the correct global inclusion probabilities.

\subsubsection{Iterator-Based Sampling Interface}

\texttt{itsample} provides a convenience layer that dispatches between stream-oriented and length-aware implementations using iterator traits. When \texttt{Base.IteratorSize(iter)} is \texttt{Base.SizeUnknown}, it selects reservoir methods that remain valid on a one-pass stream. When the iterable has a known length, it can instead switch to sequential methods that exploit that additional information. Users who need exact control over the selected algorithm or who need to update the sample incrementally should instantiate \texttt{ReservoirSampler} or \texttt{SequentialSampler} directly.

\subsection{Example Usage}

The following minimal examples illustrate typical usage of the interface. Notice that, in all cases, the iterator is never materialized in memory:

\lstset{numbers=left, numberstyle=\tiny\color{gray}, numbersep=6pt,
        caption=\relax, captionpos=b}

\begin{lstlisting}[
language = Julia,
numbers=none,
label={lst:reservoir},
caption={Reservoir sampling without replacement of $K=10$ elements from an iterator of unknown size.}
]
using StreamSampling

stream = 1:10^8
K = 10
alg = AlgL()
sampler = ReservoirSampler{Int}(K, alg)
for x in stream
    fit!(sampler, x) # O(1) per element
end

# returns a vector of length 10
sample = value(sampler)
\end{lstlisting}

\vspace{-15pt}

\begin{lstlisting}[
language = Julia,
numbers=none,
label={lst:sequential},
caption={Sequential sampling without replacement of $K=10$ elements when the population size $N$ is known.}
]
using StreamSampling

stream = 1:10^8
K, N = 10, 10^8
alg = AlgD()
sampler = SequentialSampler{Int}(stream, K, N, alg)
for x in sampler
    println(x) # sampled element emitted on the fly
end
\end{lstlisting}

\vspace{-15pt}

\begin{lstlisting}[
language = Julia,
numbers=none,
label={lst:itsample},
caption={Sampling without replacement of $K=10$ elements with the \texttt{itsample} convenience layer.}
]
using StreamSampling

# returns a vector of length 10
itsample(1:10^8, 10)
\end{lstlisting}

The reservoir sampler maintains a buffer of exactly $K$ elements throughout and is valid even when $N$ is unknown. The sequential sampler holds no buffer at all; it skips directly to each selected element using the precomputed population size $N$.

\section{Benchmarks}
\label{sec:benchmarks}

Several benchmarks were conducted to evaluate the performance of the reservoir and sequential sampling implementations\footnotemark{}. All benchmarks were run on a machine with an AMD Ryzen 5 5600H (6 cores) and 16GB of RAM, running Ubuntu 24.04 LTS and Julia 1.12.

To evaluate the algorithms' core computational performance, we benchmarked them on a simple iterator, drawing samples of sizes ranging from $0.01\%$ to $10\%$ of a generator producing integers between 1 and $10^8$. In this setup, the baseline population method uses \texttt{StatsBase.sample}, which requires fully materializing the iterator into an array before sampling. This imposes a significant memory footprint and allocation overhead.  

The reservoir and sequential methods provided by the package bypass collecting the iterator into memory entirely, drastically reducing allocations. While reservoir algorithms naturally handle iterators of unknown length in a single pass, sequential sampling algorithms require the total weight or population size to be known in advance to calculate inclusion probabilities. If this total is unknown, a prior pass over the iterator is required, so a two-passes method is also included in the comparison.

Figure \ref{fig:bench_iter} summarizes the results of the benchmarks. Each benchmark was run for at least 20 seconds of wall time, recording the median execution time and memory allocations across four sampling scenarios: unweighted/weighted and with/without replacement:

\vspace{-10pt}
\begin{figure}[h]
\centering
\includegraphics[width=\columnwidth]{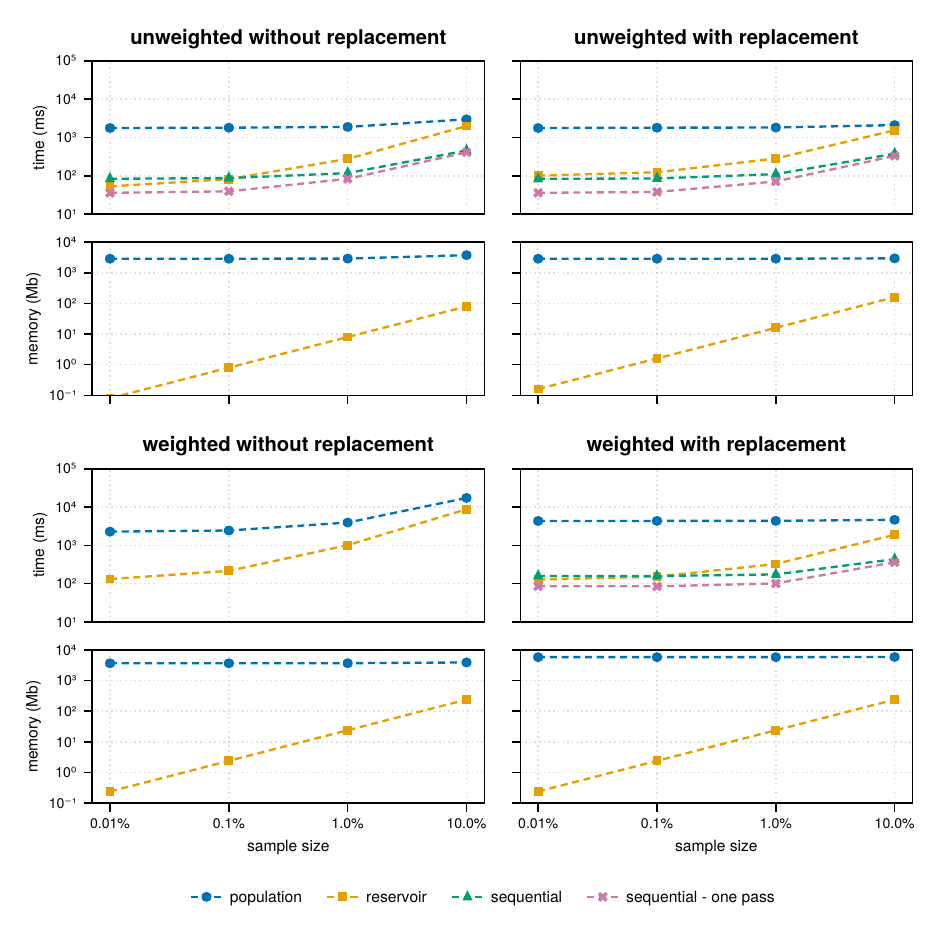}
\caption{Comparison of median execution time (ms) and memory allocation (MB) for sampling varying sizes ($10^4$ to $10^7$ elements) from an iterator of $10^8$ elements.}
\label{fig:bench_iter}
\end{figure}
\vspace{-5pt}

The population baseline forces full iterator materialization, resulting in the highest memory usage, while no allocation is performed by sequential algorithms so no line is shown for those. The reservoir methods can be seen to operate instead within $\mathcal{O}(K)$ memory. As expected by previous discussions, reservoir and sequential methods are much faster than population-based methods for small sample sizes. The one pass sequential variation demonstrates the time savings achieved when the total population weight is known in advance, allowing the algorithm to avoid a secondary traversal of the iterator.

\footnotetext{The benchmarks and applications described are available at \url{https://github.com/JuliaDynamics/StreamSampling.jl/tree/main/benchmark}.}

\section{Applications}
\label{sec:applications}

The algorithms provided by \texttt{StreamSampling.jl} can be applied across various data-intensive domains including approximate query processing, online log monitoring, querying massive network analytics feeds, and sensor data sub-sampling where unbounded observations naturally prohibit offline computational constraints \cite{cormode2011, wolfrath2022}.

A direct application is efficient sampling from persistent data\footnotemark[\value{footnote}]. To demonstrate the usefulness of the algorithms presented earlier in this setting, we performed an experiment to sample from 100 GB of weighted tabular data stored on disk in the Arrow format (a columnar binary format enabling efficient I/O without full deserialization). As a baseline, a chunking approach which loads and samples each chunk with \texttt{StatsBase.sample} and eventually recombines the partial samples is provided. Reservoir techniques process the elements in a continuous single pass which can allow them to outperform the chunking approach. Sequential methods are similarly applicable; however, they require two passes, one to compute the aggregate total weight $W_N$, and a second pass to extract the sample.

\vspace{-15pt}
\begin{figure}[h]
\centering
\includegraphics[width=\columnwidth]{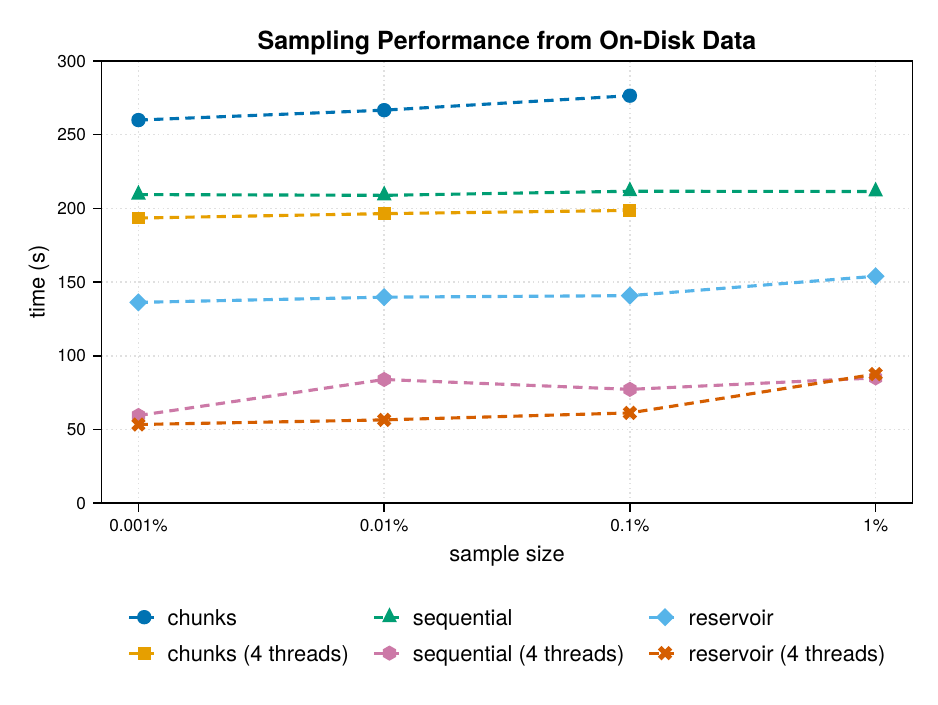}
\caption{Wall-clock time for weighted sampling with replacement from 100 GB of Arrow tabular data on disk, as a function of sample size $K$ (shown as a percentage of the total row count $N \approx 3.1 \times 10^9$). The chunks baseline crashed at the largest sample size.}
\label{fig:bench_disk}
\end{figure}
\vspace{-5pt}

The out-of-core experiment was conducted on the same machine and in the same environment described earlier for the iterator benchmarks. In addition, it should be noted that a 512 GB KIOXIA KXG60ZNV SSD was used as persistent storage. File I/O was performed via \texttt{Arrow.jl} and OS page-cache effects were mitigated by dropping the page cache between runs. Execution times are single-trial measurements.

Figure \ref{fig:bench_disk} presents execution times for extracting a weighted sample with replacement from the 100 GB Arrow file with the different methods.

The baseline chunked \texttt{StatsBase.sample} strategy exhibits the highest execution times across all tested configurations. The single-threaded chunks approach requires between 260 and 280 seconds, while the 4-thread version improves this to approximately 195 seconds. Furthermore, both chunk configurations fail due to out-of-memory errors before completing the 1\% sample size extraction.

In contrast, the reservoir and sequential methods (which use \texttt{AlgWRSWRSKIP} and \texttt{AlgORDWSWR} respectively) successfully process the largest sample sizes while requiring substantially lower execution times. For single-threaded execution, the reservoir method is the most efficient, operating in the 135–155 second range and outperforming the single-threaded sequential method, which remains flat at approximately 210 seconds. However, the sequential method demonstrates superior parallel scaling in comparison to the reservoir method.

\section{Conclusion}

\texttt{StreamSampling.jl} introduces a comprehensive suite of algorithms covering both reservoir and sequential paradigms, filling a gap in the Julia ecosystem. The ability to sample from any Julia iterator without materializing it in memory extends the reach of random sampling to contexts such as lazy data pipelines and large on-disk datasets. Future work may aim to extend the library to support other sampling algorithms. This includes implementing methods where the sample size is only guaranteed in expectation, such as Bernoulli and Poisson sampling. Furthermore, we plan to introduce sliding window sampling techniques \cite{braverman2012optimal} as well as stratified reservoir algorithms \cite{lee2010stratified}. Finally, we aim to integrate the package algorithms into the broader Julia ecosystem, specifically by improving the poly-algorithm implementation of \texttt{StatsBase.sample} to dispatch to these methods under favorable circumstances.

\section{AI Usage Disclosure}

During the preparation of this work, the author used some AI tools for the purpose of proofreading and improving readability. On the software side, these have also been used to add tests to improve code coverage. The author reviewed and edited the content as needed and takes full responsibility for the final content of the publication.

\balance
\bibliographystyle{juliacon}
\bibliography{ref}

\end{document}

%% file: header.tex

\title{StreamSampling.jl: Efficient Sampling from Data Streams in Julia}

\author[1*]{Adriano Meligrana}
\affil[1]{Sapienza University of Rome}

\keywords{Julia, Reservoir Sampling, Sequential Sampling, Online Sampling, Data Streams}

\hypersetup{
pdftitle = {StreamSampling.jl: Efficient Sampling from Data Streams in Julia},
pdfsubject = {JuliaCon 2026 Proceedings},
pdfauthor = {Adriano Meligrana},
pdfkeywords = {Julia, Reservoir Sampling, Sequential Sampling, Online Sampling, Data
Streams},
}